\newcommand{\myname}{Geoff Boeing}
\newcommand{\myemail}{g.boeing@northeastern.edu}
\newcommand{\myaffiliation}{School of Public Policy and Urban Affairs\\Northeastern University}
\newcommand{\paperdate}{November 2018}
\newcommand{\papertitle}{Planarity and Street Network Representation in Urban Form Analysis}
\newcommand{\papercitation}{Boeing, G. 2018. \papertitle. \textit{Environment and Planning B: Urban Analytics and City Science}. doi:10.1177/2399808318802941}
\newcommand{\paperkeywords}{GIS, OpenStreetMap, planar graphs, street networks, urban form, urban morphology}

\RequirePackage[l2tabu,orthodox]{nag}   
\documentclass[11pt,twocolumn]{article} 

\usepackage[T1]{fontenc}                
\usepackage[utf8]{inputenc}             
\usepackage{crimson}                    
\usepackage{helvet}                     

\usepackage[strict,autostyle]{csquotes} 
\usepackage[USenglish]{babel}           
\usepackage{microtype}                  

\usepackage{abstract}                   
\usepackage{authblk}                    
\usepackage{booktabs}                   
\usepackage{caption}                    
\usepackage[final]{draftwatermark}      
\usepackage{endnotes}                   
\usepackage{geometry}                   
\usepackage{graphicx}                   
\usepackage{hyperref}                   
\usepackage{natbib}                     
\usepackage{rotating}                   
\usepackage{setspace}                   
\usepackage{titlesec}                   
\usepackage{url}                        

\graphicspath{{./figures/}}

\titleformat{\section}{\normalfont\sffamily\large\bfseries\color{black}}{\thesection.}{0.3em}{}
\titleformat{\subsection}{\normalfont\sffamily\small\bfseries\color{black}}{\thesubsection.}{0.3em}{}

\hypersetup{
	pdfauthor={\myname},
	pdftitle={\papertitle},
	pdfsubject={\papertitle},
	pdfkeywords={\paperkeywords},
	pdffitwindow=true,         
	breaklinks=true,           
	hidelinks=true             
}

\SetWatermarkText{DRAFT}
\SetWatermarkScale{1.3}
\SetWatermarkLightness{0.9}

\begin{document}

\title{\papertitle\footnote{{Cite as: \papercitation}}}
\date{\paperdate}
\author[]{\myname\footnote{Email: \href{mailto:\myemail}{\myemail}}}
\affil[]{\myaffiliation}

\twocolumn[
\maketitle
\begin{onecolabstract}
Models of street networks underlie research in urban travel behavior, accessibility, design patterns, and morphology. These models are commonly defined as planar, meaning they can be represented in two dimensions without any underpasses or overpasses. However, real-world urban street networks exist in three-dimensional space and frequently feature grade separation such as bridges and tunnels: planar simplifications can be useful but they also impact the results of real-world street network analysis. This study measures the nonplanarity of drivable and walkable street networks in the centers of 50 cities worldwide, then examines the variation of nonplanarity across a single city. It develops two new indicators---the Spatial Planarity Ratio and the Edge Length Ratio---to measure planarity and describe infrastructure and urbanization. While some street networks are approximately planar, we empirically quantify how planar models can inconsistently but drastically misrepresent intersection density, street lengths, routing, and connectivity.
\vspace{1cm}
\end{onecolabstract}]
\saythanks 

\section{Introduction}

In urban planning and transportation research, street networks are routinely used to calculate accessibility between origins and destinations or to compute indicators of the urban form such as block sizes, intersection density, and connectivity. Mathematical models of street networks, called \emph{graphs}, have grown ubiquitous in the urban studies literature in recent years as they have been used to model household travel behavior, access and equity, pedestrian volume, urban design patterns, spatial morphology, and location centrality and polycentricity \citep{marshall_street_2010,pflieger_switzerland_2010,porta_street_2012,lee_identifying_2014,porta_alterations_2014,marshall_community_2014,hajrasouliha_impact_2015,parthasarathi_street_2015,knight_metrics_2015,xiao_identifying_2016,zhong_revealing_2017}.

The urban studies literature often defines street networks as \enquote{planar} or \enquote{approximately planar,} meaning they can be well-represented by a simplified two-dimensional model that inherently precludes overpasses or underpasses. However, the analytical impacts of this computationally useful simplification have not been studied empirically. In the real world, street networks are embedded in three-dimensional space and often feature grade separation, bridges, and tunnels. How well do planar graphs model urban street networks? To what extent do convenient planar simplifications alter network analysis results? What does the extent to which a network is nonplanar tell us about a city's infrastructure and development?

This study tests the planarity of street networks around the world and demonstrates how planar simplifications affect the results of street network analysis. It also presents two new measures of the \enquote{extent of nonplanarity} that describe the urban form and transportation infrastructure and can be generalized to other types of spatial networks. This study empirically quantifies how planarity misrepresents the street networks of many urban centers: while some street networks are well-modeled by planar graphs, others are substantially misrepresented and the magnitude of this bias varies across cities and urbanization types.

This paper first introduces the basics of graph theory relevant to urban studies, focusing on discussions in the research literature about street network planarity. Next it discusses the methods used to acquire and analyze street networks in this study. Then it presents the results of this analysis before concluding with a discussion of implications for street network research and urban form studies.

\section{Background and motivation}

\subsection{Street network models}

Graph theory is the mathematical study of networks \citep{newman_networks:_2010}. Graphs can model real-world networks such as friendships, the world wide web, or spatial networks such as urban street networks \citep{barthelemy_spatial_2011}. A graph $G$ consists of a set of nodes $N$ connected to one another by a set of edges $E$. An edge $uv$ in a directed graph points in one direction from some node $u$ to some node $v$, while an undirected graph's edges all point mutually in both directions. In a street network, the nodes represent intersections and dead-ends, and the (directed) edges represent the street segments that connect them. How a graph's nodes and edges connect to one another defines its \emph{topology}. For example, a node's \emph{degree} is the topological trait that represents how many edges connect to that node.

A \emph{planar} graph can be drawn on a two-dimensional plane without any of its edges crossing each other, except where they intersect at nodes. If it cannot be drawn to meet this criterion, it is nonplanar \citep{trudeau_introduction_1994}. Street networks are embedded in space, which provides them with geometry---such as geographical coordinates, lengths, areas, shapes, and angles---along with their topology. This creates a minor wrinkle when we consider planarity: we must distinguish between a graph's mathematical/topological planarity, which we refer to as \emph{formal planarity}, and the planarity of its real-world spatial embedding, which we refer to as \emph{spatial planarity}. For example, a street network might be spatially nonplanar due to its embedding in space (i.e., it contains overpasses or underpasses in the real world), but it could still be formally planar. That is, if we \enquote{redraw} the graph by moving its nodes and edges around in space without changing how they are connected to one another (i.e., altering its geometry without altering its topology), there may exist some alternative spatial embedding that prevents edges crossing anywhere but at nodes \citep[for a thorough introduction see][p.~6]{barthelemy_morphogenesis_2017}. In such a case, the street network is formally planar from a topological perspective, but its real-world embedding is spatially nonplanar.

\begin{figure}[tbp]
	\centering
	\includegraphics[width=0.48\textwidth]{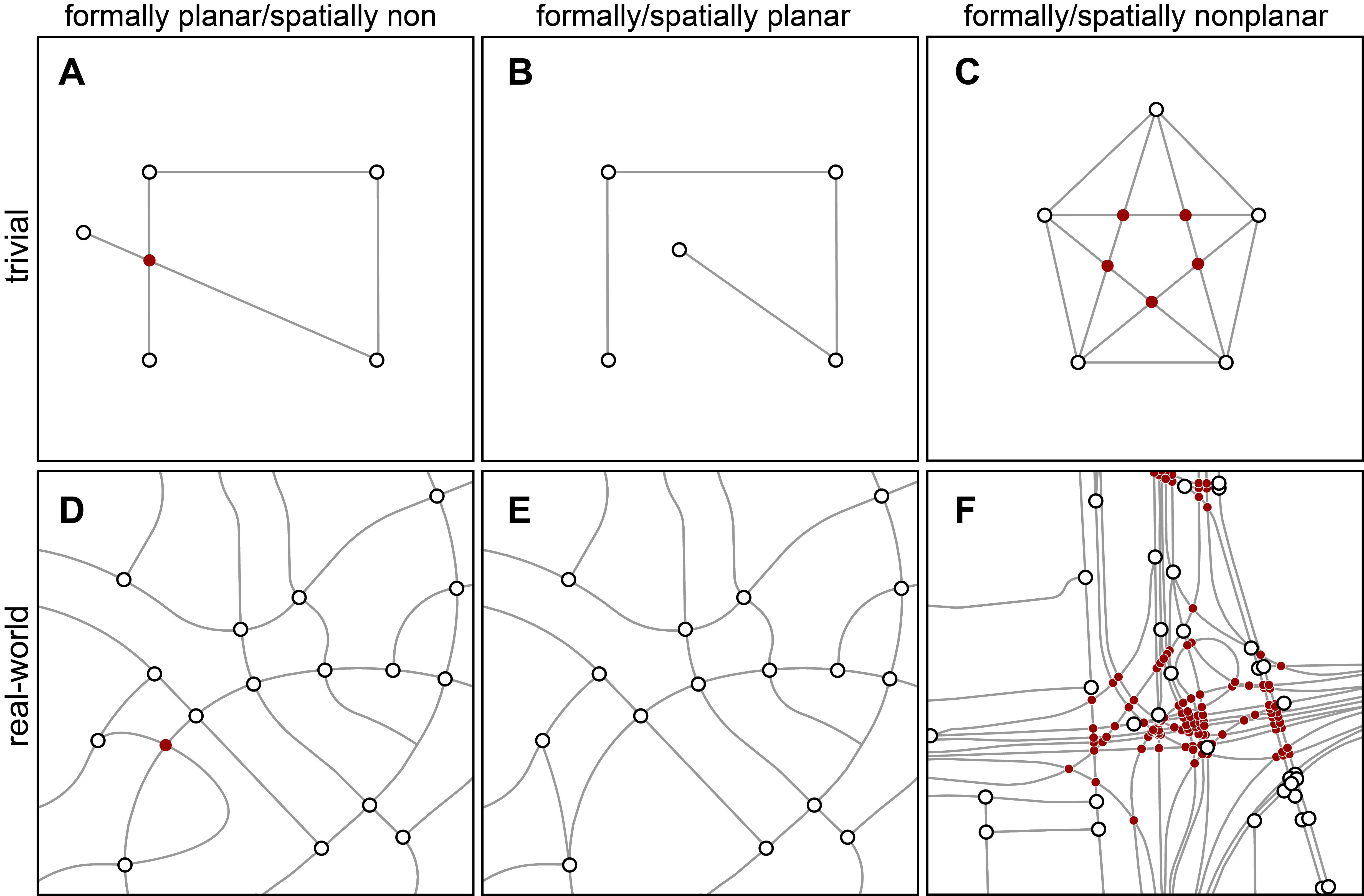}
	\caption{Spatial network examples. Gray lines = edges (i.e., street segments), circles = nodes (i.e., intersections and dead ends), red dots = two-dimensional line crossings (i.e., artificial nodes in planar representation).}
	\label{fig:planar_vs_not}
\end{figure}

Figure \ref{fig:planar_vs_not}'s top row illustrates trivial examples of this. \ref{fig:planar_vs_not}A depicts a spatially nonplanar network (with one non-node line crossing represented by a red dot) that is formally planar since it can be \enquote{redrawn} (\ref{fig:planar_vs_not}B) such that its edges only intersect at nodes. By contrast, \ref{fig:planar_vs_not}C is both formally and spatially nonplanar because it cannot be redrawn to prevent line crossings. Figure \ref{fig:planar_vs_not}'s bottom row depicts real-world examples. \ref{fig:planar_vs_not}D's road network contains a single overpass (red dot) and is thus spatially nonplanar. However, it is formally planar as we can redraw it (\ref{fig:planar_vs_not}E), by rearranging an edge in space without breaking any connections, so its edges now intersect only at nodes. The network in \ref{fig:planar_vs_not}F includes a freeway interchange with overpasses and underpasses. In two dimensions, its edges frequently cross each other at non-nodes, so it is spatially nonplanar. Furthermore, it is mathematically impossible to redraw the graph such that its edges only intersect at nodes. Therefore it is also formally nonplanar.

Imposing a planar model on a street network forces artificial nodes at any line crossings. At a citywide or regional spatial scale, these nonplanar edge crossings may be relatively uncommon: we might call such a network \emph{approximately} planar. Approximate planarity constrains nonplanar spatial networks such that they do not exhibit certain characteristics found among nonplanar aspatial graphs, such as small-world effects or power-law distributed node degrees \citep{crucitti_centrality_2006,fischer_spatial_2014}.

\subsection{Planar models in urban research}

In the urban studies literature, street networks are commonly modeled as planar graphs. Many street networks are obviously nonplanar in reality, but researchers use planar models as a potentially useful simplification of real-world complexity. Many scholars argue that street networks are planar graphs (e.g., \citealp[p.~18]{batty_network_2005}; \citealp[p.~521]{buhl_topological_2006}; \citealp[p.~1]{hu_topological_2008}; \citealp[p.~259]{masucci_random_2009}; \citealp[p.~114]{porta_networks_2010}; \citealp[p.~3]{strano_elementary_2012}; \citealp[p.~1]{masucci_limited_2013}; \citealp[p.~1074]{strano_urban_2013}; \citealp[p.~168]{law_defining_2017}). Others prefer to hedge somewhat, arguing that street networks are approximately or essentially planar---close enough to be well-modeled as such (e.g., \citealp[p.~6]{dill_measuring_2004}; \citealp[p.~3]{cardillo_structural_2006};  \citealp[p.~340]{xie_measuring_2007}; \citealp[p.~1]{barthelemy_modeling_2008}; \citealp[p.~3]{barthelemy_spatial_2011}; \citealp[pp.~563]{chan_urban_2011}; \citealp[p.~1]{gudmundsson_entropy_2013}; \citealp[p.~1]{viana_simplicity_2013}; \citealp[p.~2]{louf_typology_2014}; \citealp[p.~2191]{zhong_detecting_2014}; \citealp[p.~2]{wang_resilience_2015}; \citealp[p.~42]{aldous_routed_2016}; \citealp[p.~257]{barthelemy_paths_2017}).

If street networks can be sufficiently well-modeled by planar graphs, there are certain methodological benefits to doing so. Planar graphs offer computational simplicity and algorithmic tractability. They enable simple polygonal analysis of city blocks and form \citep{fohl_non-planar_1996,barthelemy_paths_2017}, the comparative study of topological differences between networks \citep{ahmed_local_2014,abdelkader_topological_2018}, and the calculation of the \emph{alpha} and \emph{gamma} indices popular in road network analysis \citep{eppstein_studying_2008}. Classifying planar graphs represents a trivial problem \citep{louf_typology_2014}. Planar graphs are easier to visualize and can be much faster to run algorithms on \citep{liebers_planarizing_2001}. Accordingly, \citet[p.~3]{barthelemy_spatial_2011} argues that \enquote{planar spatial networks are the most important and most studies have focused on these examples}. In contrast, \citet{masucci_random_2009} and \citet{masucci_limited_2013} argue that planar graphs remain a compelling research domain for urban scholars because they were understudied until recently for two reasons: they appear topologically trivial and planarity does not lend itself to certain popular analyses of nonplanar aspatial graphs.

Although planar models can be computationally useful, real-world street networks often include at least one overpass or underpass that results in the failure of formal proofs of planarity, such as the \citet{kuratowski_sur_1930} theorem or the \cite{hopcroft_efficient_1974} algorithm \citep[cf.][]{gastner_spatial_2006,levinson_network_2012}. \citet[p.~199]{jiang_object-oriented_2010} note that \enquote{quite often the transportation network has overpasses and underpasses that require a non-planar network representation.} \citet[p.~1258]{fischer_spatial_2014} explains that \enquote{for many infrastructure networks, {[planarity]} is approximately true, although bridges and tunnels in ground-transport networks are an obvious (but generally minor) exception.} However, the presence of such nonplanar elements can vex models. \enquote{Despite its popularity, the [planar] model has limitations for some areas of transportation analysis, especially where complex network structures are involved. One major problem is caused by the planar embedding requirement... intersections at grade cannot be distinguished from intersections with an overpass or underpass that do not cross at grade} \citep[p.~395]{fischer_gis_2004}.

If a planar graph models a street network poorly, it could do so in multiple ways. Imposing planarity on a nonplanar street network produces artificial nodes in the graph at overpasses and underpasses, which breaks routing. For example, this misrepresentation would allow a routing engine to recommend a left turn at an \enquote{intersection} that is actually an overpass \citep[p.~6]{kwan_review_1996}. Intersection counts and densities would be overestimated by the presence of these false nodes, and edge lengths would be underestimated due to these artificial breakpoints splitting up street segments. This consequently impacts interpretation of urban form density, grain, pattern, connectedness, and permeability. Finally, this bias might behave inconsistently across different kinds of cities and street network types based on the extents to which they are planar.

\subsection{Open questions}

Given these issues, some unanswered questions remain in the empirical literature. What do \emph{approximately planar} and \emph{well-modeled} mean for street network research? How do planar simplifications impact the results of street network analyses? Do the biases of planar models behave consistently across geographies and development eras or do they misrepresent different cities (or neighborhoods) to different extents? And if street networks are at least generally approximately planar, how can we measure the extent to which a given street network is nonplanar?

The graph theory literature offers some measures of how \enquote{far off} a nonplanar graph $G$ is from being planar, including its \emph{crossing number}---the minimum number of edge crossings of any drawing of $G$---and its \emph{skewness}---the minimum number of edges that must be removed from $G$ to produce a planar graph \citep{liebers_planarizing_2001,chimani_non-planar_2009}. However, these measures are imperfect and hard to compute \citep{szekely_successful_2004,chimani_vertex_2012}. They also fail to adjust for the size, density, or real-world embedding of a spatial network. Discussing road networks and approximate spatial planarity, \citet[p.~133]{newman_networks:_2010} argues that due to such drawbacks no widely-accepted measure of the extent of nonplanarity has emerged, and calls for the development of better indicators.

Such measures would be particularly useful for street networks, as the extent to which a network is (or is not) planar can characterize the nature of its circulation infrastructure and urban form. For instance, wealthy 20th- and 21st-century cities organized around automobility and freeways might exhibit more nonplanarity than do older cities or informal settlements oriented around walking. Beyond the question of graph model goodness-of-fit, such indicators could provide useful information about urban development, civil infrastructure, and transportation system character.

\section{Methods}

This study builds on this rich body of street network research by developing two new measures of the extent to which a spatial network is planar. It empirically analyzes various city centers worldwide to quantify how planar models impact street network analysis results (focusing on the common urban form measures of intersection density and street segment length) and the extent to which bias (i.e., model misrepresentation) varies across places and types of urbanization. Finally, it considers what these indicators suggest about the urban form and transportation infrastructure in different cities.

\subsection{Data}

Following \citet{jacobs_great_1995} and \citet{cardillo_structural_2006}, we analyze a consistently sized, square-mile network at the centers of 50 major cities worldwide. This allows us to consistently examine central urban street networks without being swamped by metropolitan-scale variation or the idiosyncrasies of individual municipalities' spatial extents. The 50 sampled cities span Africa, Asia, Australia, Europe, and North and South America. We look separately at each city's drivable and walkable street networks. For cities with a newer central business district (CBD) that lies apart from an \enquote{old town} center, we take the modern CBD as the city center.

To acquire these street networks, we use OSMnx to download the data for each city and network type from OpenStreetMap. OpenStreetMap is a collaborative online mapping platform commonly used by researchers because of its high-quality worldwide coverage \citep{haklay_how_2010,jokar_arsanjani_openstreetmap_2015}. OSMnx is a Python-based software tool that allows us to download street network data from OpenStreetMap for any study site in the world, then process it into a length-weighted nonplanar directed graph \citep{boeing_osmnx:_2017,boeing_multi-scale_2018}. It differentiates between walkable and drivable routes based on individual elements' metadata that describe how the route may be used. Thus the walkable network may contain surface streets, paths through parks, pedestrian flyovers, passageways between buildings, and other walkable paths. The drivable network may contain surface streets, grade-separated freeways, and other drivable routes.

OpenStreetMap's raw data contain many interstitial nodes in the middle of street segments (forming an expansion graph) to allow streets to curve through space via a series of straight-line approximations. OSMnx automatically simplifies each graph's topology to retain nodes only at intersections and dead-ends, while retaining each edge's true spatial geometry. This provides accurate intersection counts and edge length measures for comparison between these networks' planar and nonplanar representations.

\subsection{Analysis}

Once we have acquired and prepared the networks, we calculate three measures of planarity. The first is an aspatial test of formal planarity using the algorithm described by \citet{boyer_subgraph_2012}. This assesses if it is possible to rearrange the graph's nodes and edges in space, while preserving its topology, so that edges cross only at nodes. This binary true/false indicator tells us if the graph is formally planar, ignoring its real-world spatial embedding. However, street networks \emph{are} spatially embedded: every spatially planar network is formally planar, but not every formally planar network is spatially planar. Accordingly, for this study, we have developed a second and third indicator to assess the \enquote{extent} to which they are planar.

The second measure is the Spatial Planarity Ratio, $\phi$. It represents the ratio of the number of nonplanar intersections $i_n$ (i.e., non-dead-end nodes in the nonplanar, three-dimensional, spatially-embedded graph) to the number of planar intersections $i_p$ (i.e., edge crossings in the planar, two-dimensional, spatially-embedded graph): 

\begin{equation}
	\label{eq:spr}
	\phi = \frac{i_n}{i_p}
\end{equation}

Thus, $\phi$ represents what proportion of the two-dimensional edge crossings in the planar graph are true intersections in the nonplanar graph and, accordingly, $i_p-i_n$ equals the number of nonplanar edge crossings (i.e., overpasses and underpasses) in the network. A spatially planar network with no overpasses or underpasses will have a $\phi$ score of 1.0, while lower values indicate the extent to which the network is planar. From this indicator, we can calculate how planar simplifications overstate intersection counts and connectivity in a street network (i.e., by what percentage the planar model overcounts intersections) as:

\begin{equation}
	\label{eq:spr_overstates}
	\frac{1}{\phi} - 1
\end{equation}

The third measure is the Edge Length Ratio, $\lambda$. It represents the ratio of the planar graph's mean edge length $l_p$ to the nonplanar graph's mean edge length $l_n$:

\begin{equation}
	\label{eq:elr}
	\lambda = \frac{l_p}{l_n}
\end{equation}

This measures how planar simplifications fragment street segments at overpasses or underpasses. A spatially planar street network will thus have a $\lambda$ score of 1.0, while lower values indicate the extent to which the network is planar. We can calculate by what percentage the planar graph underestimates the average edge length as:

\begin{equation}
	\label{eq:elr_understates}
	1 - \lambda
\end{equation}

Thus, equations \ref{eq:spr} and \ref{eq:elr} measure the extent to which a spatial network is planar, while equations \ref{eq:spr_overstates} and \ref{eq:elr_understates} measure the extent to which a planar model misrepresents the network's characteristics.

Finally, street network models almost always impose an artificial boundary on the network, and moving the study site even slightly might affect these results \citep{gil_street_2017}. We thus explore how these three indicators vary throughout a single city by analyzing the drivable street network of Oakland, California as a case study. Oakland is a mid-sized American city with a variety of urban form types from gridded street patterns in its flatlands, to winding culs-de-sac in its hills, to freeways and dense blocks around its downtown. First we analyze the entire city of Oakland. Then we randomly sample 100 points within the city limits and analyze the square-mile street networks centered on each. The resulting statistical dispersion of planarity demonstrates the extent to which analyzing an entire city's neighborhoods as a single graph may obscure neighborhood-scale infrastructure characteristics.

\section{Results}

\begin{table*}[htbp]
	\centering
	\footnotesize
	\caption{Planarity measures for central street networks in 50 cities worldwide (Planar = whether street network passed the formal test of planarity; $\phi$ = Spatial Planarity Ratio; $\lambda$ = Edge Length Ratio).}
	\label{tab:world_cities}
	\begin{tabular}{ l l r r r r r r  }
\toprule
         &               & \multicolumn{3}{|c|}{Drive}         & \multicolumn{3}{c}{Walk}            \\
\midrule
Country      & City          &  Planar  &  $\phi$   &  $\lambda$   &  Planar  &  $\phi$   &  $\lambda$   \\
	\midrule
	Argentina & Buenos Aires &      Yes &  1.000 &  1.000 &       No &  0.939 &  0.941 \\
	Australia & Sydney &       No &  0.729 &  0.735 &       No &  0.902 &  0.884 \\
	Brazil & Sao Paulo &       No &  0.771 &  0.772 &       No &  0.824 &  0.803 \\
	Canada & Toronto &      Yes &  0.922 &  0.946 &       No &  0.838 &  0.824 \\
	& Vancouver &       No &  0.930 &  0.948 &       No &  0.923 &  0.920 \\
	Chile & Santiago &       No &  0.873 &  0.885 &       No &  0.967 &  0.965 \\
	China & Beijing &       No &  0.818 &  0.846 &       No &  0.842 &  0.842 \\
	& Hong Kong &       No &  0.838 &  0.823 &       No &  0.823 &  0.794 \\
	& Shanghai &       No &  0.682 &  0.708 &       No &  0.660 &  0.641 \\
	Denmark & Copenhagen &      Yes &  0.992 &  0.988 &       No &  0.994 &  0.985 \\
	Egypt & Cairo &       No &  0.897 &  0.913 &       No &  0.899 &  0.886 \\
	France & Lyon &       No &  0.995 &  0.991 &       No &  0.958 &  0.953 \\
	& Paris &       No &  0.982 &  0.987 &       No &  0.928 &  0.907 \\
	Germany & Berlin &       No &  0.939 &  0.945 &       No &  0.939 &  0.931 \\
	India & Delhi &      Yes &  1.000 &  1.000 &      Yes &  0.997 &  0.989 \\
	Indonesia & Jakarta &      Yes &  0.990 &  0.994 &       No &  0.969 &  0.965 \\
	Iran & Tehran &       No &  0.962 &  0.973 &       No &  0.953 &  0.951 \\
	Italy & Bologna &      Yes &  1.000 &  1.000 &      Yes &  0.996 &  0.996 \\
	& Florence &      Yes &  0.993 &  0.994 &       No &  0.980 &  0.974 \\
	& Milan &      Yes &  1.000 &  1.000 &       No &  0.849 &  0.832 \\
	Japan & Osaka &       No &  0.868 &  0.868 &       No &  0.953 &  0.950 \\
	& Tokyo &       No &  0.925 &  0.919 &       No &  0.924 &  0.912 \\
	Kenya & Nairobi &       No &  0.974 &  0.971 &       No &  0.949 &  0.938 \\
	Mexico & Mexico City &       No &  0.940 &  0.949 &       No &  0.912 &  0.917 \\
	Nigeria & Lagos &       No &  0.960 &  0.972 &       No &  0.990 &  0.988 \\
	Peru & Lima &       No &  0.941 &  0.951 &       No &  0.923 &  0.921 \\
	Philippines & Manila &       No &  0.940 &  0.947 &       No &  0.897 &  0.883 \\
	Russia & Moscow &       No &  0.540 &  0.596 &       No &  0.842 &  0.833 \\
	Singapore & Singapore &       No &  0.864 &  0.869 &       No &  0.891 &  0.880 \\
	Somalia & Mogadishu &      Yes &  1.000 &  1.000 &      Yes &  1.000 &  1.000 \\
	South Africa & Johannesburg &       No &  0.847 &  0.877 &       No &  0.997 &  0.997 \\
	Spain & Barcelona &      Yes &  1.000 &  1.000 &       No &  0.925 &  0.917 \\
	Switzerland & Geneva &       No &  0.985 &  0.981 &       No &  0.834 &  0.807 \\
	Thailand & Bangkok &       No &  0.993 &  0.979 &       No &  0.990 &  0.984 \\
	Turkey & Istanbul &       No &  0.965 &  0.964 &       No &  0.973 &  0.964 \\
	UAE & Dubai &       No &  0.679 &  0.668 &       No &  0.852 &  0.837 \\
	UK & Edinburgh &       No &  0.974 &  0.965 &       No &  0.987 &  0.983 \\
	& London &       No &  0.976 &  0.980 &       No &  0.853 &  0.836 \\
	USA & Atlanta &       No &  0.720 &  0.765 &       No &  0.805 &  0.788 \\
	& Chicago &       No &  0.748 &  0.786 &       No &  0.792 &  0.787 \\
	& Cincinnati &       No &  0.723 &  0.746 &       No &  0.929 &  0.922 \\
	& Dallas &       No &  0.584 &  0.639 &      Yes &  0.961 &  0.956 \\
	& Los Angeles &       No &  0.581 &  0.627 &       No &  0.784 &  0.785 \\
	& Miami &       No &  0.641 &  0.657 &       No &  0.962 &  0.961 \\
	& New York &       No &  0.879 &  0.878 &       No &  0.928 &  0.923 \\
	& Phoenix &       No &  0.949 &  0.958 &       No &  0.977 &  0.972 \\
	& San Francisco &       No &  0.935 &  0.937 &       No &  0.942 &  0.935 \\
	& Seattle &       No &  0.728 &  0.771 &       No &  0.931 &  0.924 \\
	& Washington DC &       No &  0.948 &  0.956 &       No &  0.964 &  0.961 \\
	Venezuela & Caracas &       No &  0.953 &  0.957 &      Yes &  1.000 &  1.000 \\
	\bottomrule
\end{tabular}
\end{table*}

\begin{figure*}[htbp]
	\centering
	\includegraphics[width=\textwidth]{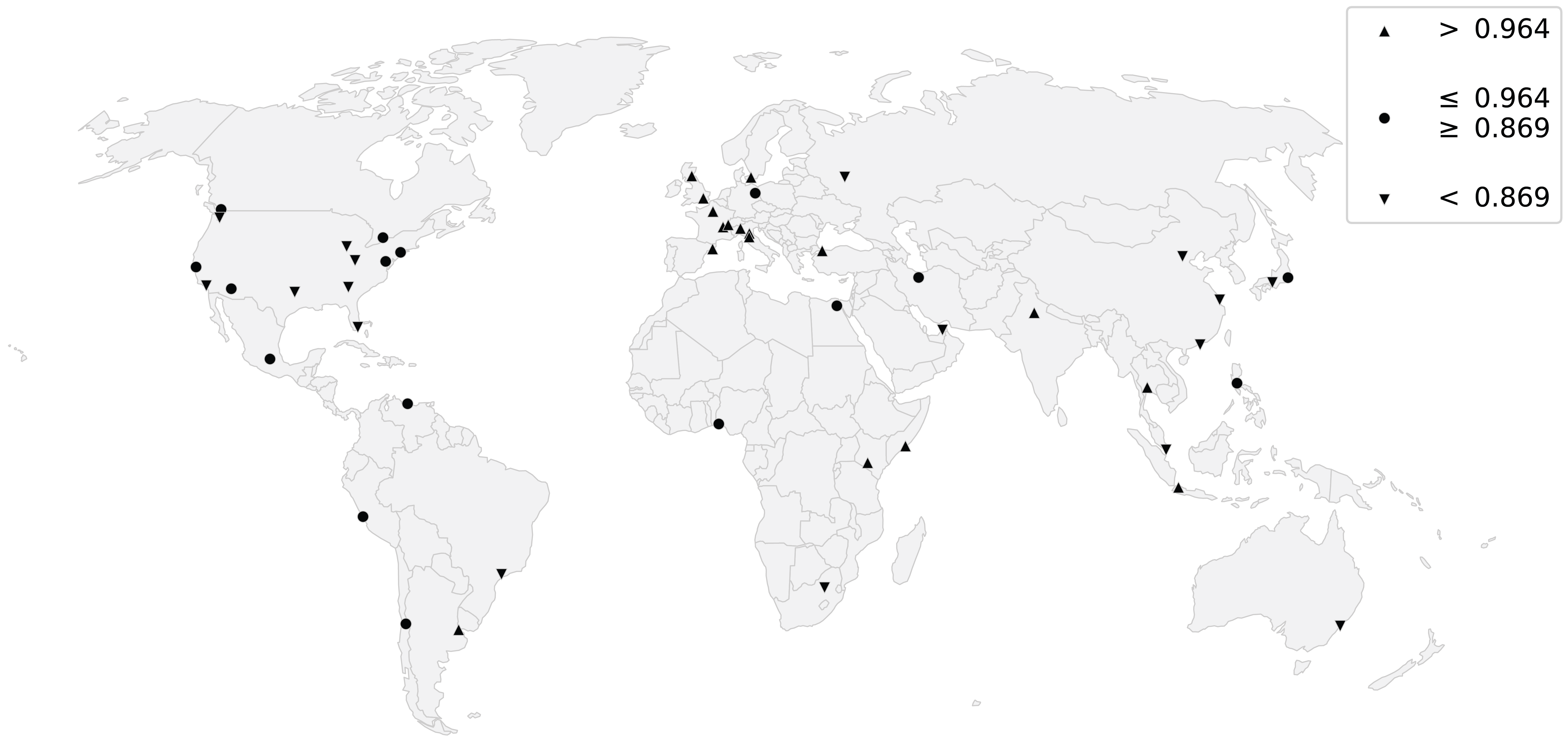}
	\caption{Map of cities from Table \ref{tab:world_cities} grouped by drivable Spatial Planarity Ratio ($\phi$) terciles (lower values mean less planar).}
	\label{fig:world_map_bw}
\end{figure*}

\begin{figure}[htbp]
	\centering
	\includegraphics[width=0.48\textwidth]{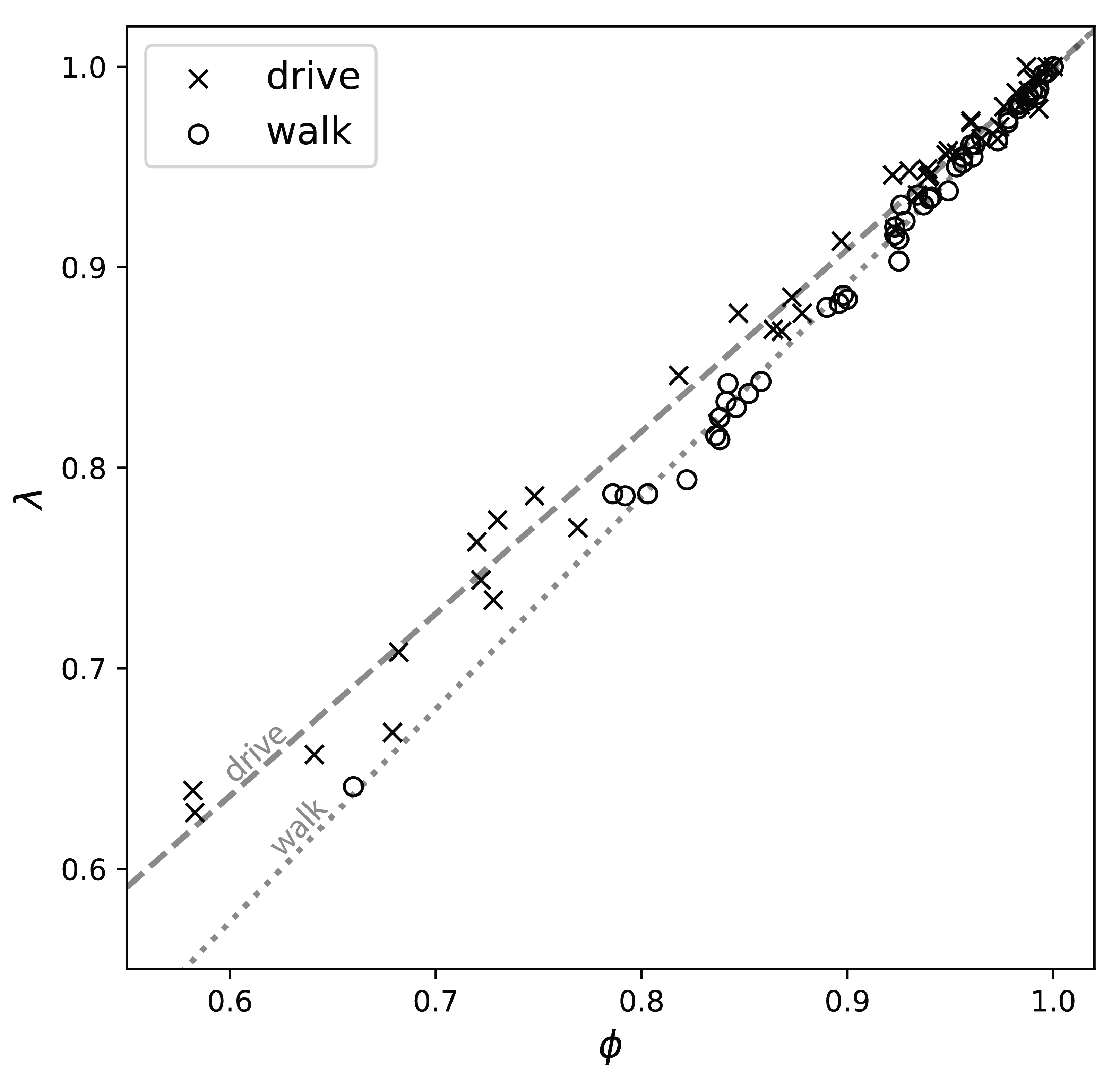}
	\caption{Scatter plot of Edge Length Ratio ($\lambda$) vs Spatial Planarity Ratio ($\phi$) by network type with regression lines.}
	\label{fig:regression_lambda_phi}
\end{figure}

Table \ref{tab:world_cities} demonstrates how planar simplifications impact street network analysis results in these 50 city centers. Among the drivable street networks, only 20\% are formally planar. On average, they are 88\% spatially planar by the $\phi$ measure (Spatial Planarity Ratio) and 89\% by the $\lambda$ measure (Edge Length Ratio). The individual $\phi$ values indicate that spatial planarity ranges from a high of 100\% in six of these cities to a low of 54\% in Moscow. The $\lambda$ values indicate that spatial planarity ranges from a high of 100\% in six of these cities to a low of 60\% in Moscow. On average across these networks, planar models overcount intersections by 16\% and underestimate street segment lengths by 11\%.

Among walkable street networks, only 10\% are formally planar. On average, they are 92\% spatially planar by the $\phi$ measure and 91\% by the $\lambda$ measure. The individual $\phi$ values indicate that spatial planarity ranges from a high of 100\% in two cities to a low of 66\% in Shanghai. The $\lambda$ values indicate that spatial planarity ranges from a high of 100\% in two cities to a low of 64\% in Shanghai. On average across these networks, planar models overcount intersections by 10\% and underestimate street segment lengths by 9\%. Fewer walkable than drivable networks are formally planar, but on average these walking networks are slightly more spatially planar than the driving networks.

Not all formally planar street networks are spatially planar. For example, Toronto's drivable network is formally planar but only 92\% ($\phi$) and 95\% ($\lambda$) spatially planar. In total, four drivable networks (Toronto, Jakarta, Florence, and Copenhagen) and three walkable networks (Dallas, Delhi, and Bologna) are formally planar but spatially nonplanar. About a third of the city centers studied demonstrate $\phi$ spatial planarity of 96\% or higher, suggesting that they are \enquote{approximately} planar. However, another third of the city centers are less than 87\% planar. Dallas, Los Angeles, and Moscow have $\phi$ values below 59\%, suggesting planar graphs poorly model these city centers. In fact, planarity overstates the intersection counts in these three networks by 71\%, 72\%, and 85\% respectively.

Mogadishu is the only city studied that demonstrates perfect planarity across all three indicators for both network types. Cities like Milan and Barcelona demonstrate perfect planarity in their centers' drivable networks, but not in their walkable networks. Further, the extent of nonplanarity is not consistent across network types: Dallas's walkable $\phi$ is 65\% greater than its drivable $\phi$, while Geneva's drivable $\phi$ is 15\% greater than its walkable $\phi$. Figure \ref{fig:world_map_bw} maps the distribution of $\phi$ values around the world. While nearly every European city is in the highest tercile, indicating their networks are more planar, most American cities are in the lowest tercile, indicating their networks are more nonplanar.

Figure \ref{fig:regression_lambda_phi} depicts the relationship between $\phi$ and $\lambda$ across all 50 cities for both network types. It is linear, positive, and strong (drivable $r^2=0.990$ and walkable $r^2=0.993$). The coefficients of determination suggest that these two indicators unsurprisingly provide redundant statistical information about the extent to which a network is spatially planar (i.e., a greater volume of artificial nodes in the planar graph will both increase the denominator of $\phi$ and decrease the numerator of $\lambda$, as these artificial breakpoints reduce the average street segment length). However, each indicator assesses different implications of this bias for measuring the urban form.

\begin{table}[htbp]
	\centering
	\caption{Summary statistics of planarity indicators across 100 square-mile samples of Oakland, California's drivable network.}
	\label{tab:samples_city}
	\begin{tabular}{lrr}
\toprule
         & $\phi$ &  $\lambda$   \\
\midrule
count    &  100   &  100 \\
mean     &  0.929 &  0.939 \\
$\sigma$ &  0.103 &  0.089 \\
min      &  0.569 &  0.567 \\
max      &  1.000 &  1.000 \\
\bottomrule
\end{tabular}
\end{table}

Finally, we examine how these measures behave across an entire city. Oakland's city-wide street network is formally nonplanar. In terms of spatial planarity, the city has a $\phi$ of 91.6\% and a $\lambda$ of 92.9\%. This suggests that the planar representation of Oakland's drivable street network overstates the number of intersections---and thus, the network's connectivity---by 9.2\% city-wide and understates the average edge length by 7.1\% city-wide. However, these indicators vary across the city (Table \ref{tab:samples_city}). The samples' mean $\phi$ and $\lambda$ scores are reasonably close to the city-wide values. However, the samples range from spatial planarity lows of 56.9\% ($\phi$) and 56.7\% ($\lambda$) up to highs of 100\%. 67\% of the samples pass the formal planarity test, but 63\% of the samples are at least somewhat spatially nonplanar (i.e., $\phi < 1$).

\section{Discussion}

Street networks at the centers of most major cities are both formally and spatially nonplanar. However, this depends on the scale of measurement: across an entire city there is likely to be at least one overpass or underpass somewhere, while individual neighborhoods or small towns may be entirely planar. The type and era of urbanization represent another factor. Old European towns or informal settlements in the Global South may contain fewer grade-separated roads---and thus would be more planar---than 20th-century American or 21st-century Chinese metropolises, a result of prevailing transportation and engineering technologies when the urban form developed, as well as local terrain, wealth, culture, and politics \citep{southworth_street_1995}.

Certain types of circulation, such as gondolas traversing Venice's canal network, could be consistently well-modeled under a planarity assumption. However, street networks worldwide are commonly nonplanar because they are embedded in three dimensions, not two. But because they are usually \enquote{mostly} planar (average drivable $\phi$ of 88\% and average walkable $\phi$ of 92\%), typically with only a few overpasses or underpasses, many could be described as \emph{approximately} planar. But formally, a graph is not planar because its edges \emph{usually} intersect only at nodes: by definition it is planar because its edges \emph{exclusively} intersect at nodes. However, as \citet{newman_networks:_2010} points out, debating the semantics of formal planarity may be missing the point---more interesting is the \emph{extent} to which a network is nonplanar. That is, how do planar simplifications impact modeling and analysis?

\subsection{Planarity and model bias}

George Box famously said, \enquote{All models are wrong but some are useful.} Even if they are not formally planar, can street networks be simplified to planar graphs and still be usefully well-modeled? Our results suggest that the answer depends on the study site and on the type of analysis. In limited circumstances---where the circulation network exhibits few (or ideally zero) underpasses, overpasses, or grade-separation---then certainly yes, especially if the study focuses on polygonal spatial analysis. But universally we cannot answer yes, especially for topological studies: most egregiously, imposing planarity on a nonplanar street network forces false nodes at underpasses and overpasses, breaking routing and network-based accessibility modeling. For this reason, nonplanar graphs have been the standard for decades in transportation engineering, real-world traffic assignment models, and routing engines.

But aside from routing, planar graphs are often used to characterize urban form and morphology. Do they provide useful models for this type of research? Again, only in certain circumstances, such as the urban form of Florence where the planar bias is arguably negligible. Meanwhile, in our central Los Angeles network, planarity overcounted intersections by 72\% and underestimated edge lengths by 37\%. This misrepresentation behaves inconsistently from place to place, as the magnitude of bias varies within and across cities and modes of urbanization. Misrepresentation is particularly pronounced around the downtowns of American cities due to the prevalence of freeways, overpasses, and underpasses.

Drivable networks are primarily affected by these features, but even walkable networks are affected by pedestrian flyovers and subways. In Florence, the walking network is both spatially and formally nonplanar due to the \textit{sottopassaggio} (pedestrian subway) near Stazione di Santa Maria Novella, its central train station. But even networks of non-freeway, non-pedestrian-only surface streets could easily become nonplanar due to tunnels in hilly neighborhoods or bridges over rivers.

We can see the impact of this bias in aggregate across our study sites: the results of common urban form analyses such as intersection counts are overstated by planar models (by 16\% on average in the drivable networks), while average street segment lengths are consequently understated (by 11\% on average in the drivable networks). These impact our understanding of the urban form's density, grain, pattern, connectedness, and permeability. To summarize, planarity varies both across cities as well as across different neighborhoods within individual cities. Modeling urban street networks as planar graphs can bias urban form analyses in several ways:

\begin{enumerate}
	\item{Intersection counts are overestimated due to false nodes where grade-separated edges cross}
	\item{Average edge lengths are underestimated}
	\item{Connectivity is misrepresented for routing, accessibility analysis, and topological studies}
\end{enumerate}

We may be able to shoehorn real-world data into a chosen model and perform a desired computation, but we lose some underlying ability to reason with it when the real world no longer obeys the model's rules and assumptions. Although planar simplifications offer algorithmic tractability and allow the polygonal analysis of urban blocks, these results demonstrate how they can model certain urban street networks poorly and bias analytical results. New tools like OSMnx can help mitigate these shortcomings by easily modeling street networks as nonplanar graphs.

\subsection{What can nonplanarity tell us?}

We might repurpose this discrepancy between planar and nonplanar models of a street network to elucidate characteristics of the urban form. For example, the $\phi$ and $\lambda$ scores of these US city centers indicate the greater three-dimensionality of their transportation infrastructure compared to that of these European city centers. During the 20\textsuperscript{th} century, US cities experienced massive investment in automobile-oriented roadways and are now laced by grade-separated freeways. In contrast, these European city centers feature less three-dimensionality as their streets and paths tend to cross at-grade. Examining how $\phi$ and $\lambda$ change over time in different cities could reveal the type and pace of urban development.

Many street networks are formally nonplanar while demonstrating $\phi$ and $\lambda$ values that are only slightly nonplanar. This approximate planarity is common across cities because of costs, technology, and politics. It is expensive to engineer an extensively three-dimensional network and a city fully consumed by overpasses and tunnels would degrade quality of life, making it politically infeasible. Future research can further explore how these indicators correlate with other measures of urbanization, development, investment, technology, politics, and era, as well as city age and size. While this study focused on measures of the urban form, particularly intersection counts and street segment lengths, future work can further explore impacts of planar simplifications on topological analyses and routing. It can also examine the spatial clustering and localization of nonplanarity and how networks might be studied piecewise as planar.

\subsection{Conclusion}

Street networks are often defined and modeled as planar graphs in the urban studies literature---despite typically being nonplanar in reality---to simplify real-world complexity and make certain analyses tractable. This study measured the formal and spatial planarity of central urban street networks empirically, demonstrating how planar simplifications inconsistently affect urban form analysis results. It developed two new measures of the extent to which spatial networks are nonplanar---the Spatial Planarity Ratio, $\phi$, and the Edge Length Ratio, $\lambda$---to characterize urbanization particularly through the transportation infrastructure's three-dimensionality. Although planar graphs are useful in some circumstances, they behave inconsistently within and across cities by misrepresenting connectivity, accessibility, routing, intersection densities, and street lengths. Individual analyses must consider these location-dependent tractability and accuracy trade-offs when choosing between planar and nonplanar models.

\section*{Acknowledgments}

The author wishes to thank Marta González, Brittany Fasy, and David O'Sullivan for their helpful comments and suggestions.

\IfFileExists{\jobname.ent}{\theendnotes}{}

\setlength{\bibsep}{0.05cm plus 0.05cm} 
\bibliographystyle{apalike}
\bibliography{references}

\end{document}